\documentclass[10pt,letter,journal]{IEEEtran}

\usepackage[english]{babel}
\usepackage[latin1]{inputenc}
\usepackage{bbm}
\usepackage{url}
\usepackage{amsmath,pifont,amssymb,amsfonts}
\usepackage{graphicx}
\usepackage{color}
\usepackage[table]{xcolor}
\usepackage{framed}
\usepackage{gensymb}
\usepackage{longtable}
\usepackage{array}
\usepackage{caption}
\usepackage{subcaption}
\usepackage{psfrag}
\usepackage{algorithm}
\usepackage{algpseudocode}
\usepackage{cite}
\usepackage{multirow}
\usepackage{balance}
\usepackage{epstopdf}
\usepackage{balance}
\usepackage{nth}

\graphicspath{{figs/}}

\newcommand{\dfn}[1]{\textit{#1}}

\begin{document}
\title{All that Glitters is not Bitcoin -- Unveiling the Centralized Nature of the BTC (IP) Network}

\author{
\IEEEauthorblockN{Sami Ben Mariem\IEEEauthorrefmark{1}\IEEEauthorrefmark{2}, Pedro Casas\IEEEauthorrefmark{1}, Matteo Romiti\IEEEauthorrefmark{1}, Benoit Donnet\IEEEauthorrefmark{2}, Rainer St\"utz\IEEEauthorrefmark{1}, Bernhard Haslhofer\IEEEauthorrefmark{1}\\}
\IEEEauthorblockA{\IEEEauthorrefmark{1}AIT Austrian Institute of Technology, \IEEEauthorrefmark{2}Universit\'e de Li\`ege}
}

\maketitle

\begin{abstract}
Blockchains are typically managed by peer-to-peer (P2P) networks providing the support and substrate to the so-called \emph{distributed ledger (DLT)}, a replicated, shared, and synchronized data structure, geographically spread across multiple nodes. The Bitcoin (BTC) blockchain is by far the most well-known DLT, used to record transactions among peers, based on the BTC digital currency. In this paper we focus on the network side of the BTC P2P network, analyzing its nodes from a purely network measurements-based approach. We present a BTC crawler able to discover and track the BTC P2P network through active measurements, and use it to analyze its main properties. Through the combined analysis of multiple snapshots of the BTC network as well as by using other publicly available data sources on the BTC network and DLT, we unveil the BTC P2P network, locate its active nodes, study their performance, and track the evolution of the network over the past two years. Among other relevant findings, we show that ($i$) the size of the BTC network has remained almost constant during the last 12 months -- since the major BTC price drop in early 2018, ($ii$) most of the BTC P2P network resides in US and EU countries, and ($iii$) despite this western network locality, most of the mining activity and corresponding revenue is controlled by major mining pools located in China. By additionally analyzing the distribution of BTC coins among independent BTC \emph{entities} (i.e., single BTC addresses or groups of BTC addresses controlled by the same actor), we also conclude that ($iv$) BTC is very far from being the decentralized and uncontrolled system it is so much advertised to be, with only 4.5\% of all the BTC entities holding about 85\% of all circulating BTC coins.
\end{abstract}

\IEEEpeerreviewmaketitle

\section{Introduction}\label{intro}

The Bitcoin (BTC) blockchain~\cite{hindawi_2018} is built on top of a decentralized peer-to-peer (P2P) network, used to propagate relevant information such as transactions between BTC entities or blockchain updates. Given its relevance and popularity, many studies have characterized the BTC blockchain from multiple different perspectives. In this paper we take a complementary analysis approach of BTC, focusing both on the network side of the BTC P2P network, as well as on the ledger side of the BTC blockchain network. Using first a purely network measurements-based approach, we characterize the BTC active nodes of the underlying P2P network, and study their main properties. Understanding the underlying P2P network characteristics of the BTC blockchain can be highly useful for multiple relevant use cases, including security, performance and anomaly detection, as well as relevant nodes discovery - i.e., identifying and locating the BTC miners. In addition, by analyzing the BTC ledger of transactions through simple clustering heuristics~\cite{semantics_2016,clustering_CBT17}, we study the distribution of BTC coins among independent BTC \emph{entities}, being an entity either a single BTC address holding BTC coins, or a group of BTC addresses controlled by the same actor.

Based on popular past studies on P2P networks~\cite{gnutella_2002,gnutella_2005}, we present a technique to discover the nodes of the BTC P2P network, based on active measurements. The discovery is done through a modified BTC client which can crawl the full set of nodes composing the BTC P2P network, and gather information regarding those nodes which are \emph{active} and \emph{reachable}. Using this crawler, we take multiple snapshots of the BTC P2P network spanning the last eight months -- including May 2019, and describe relevant findings regarding its active nodes. We complement the study by processing other sources of data publicly available, using in particular data available through the public BTC blockchain ledger -- relying on APIs on-line available at \url{https://www.btc.com}, \url{https://www.blockchain.com}, and \url{https://bitnodes.earn.com}.

Our combined study clearly shows that despite the so much advertised decentralized and uncontrolled nature and underlying paradigm of BTC, the system is a highly concentrated one, in every sense: in terms of \textbf{network infrastructure}, 36\% of the nodes are hosted by only 5 major cloud providers in EU and US, and 50\% of the nodes reside in only 3 countries; in terms of \textbf{mining activity}, almost 50\% of all BTC blocks are mined by 4 major mining pools in China; at last, the Gini index -- i.e., a typical measure of income distribution --, of BTC coins distribution among independent entities is extremely high, i.e., above 0.98, with only 4.5\% of all BTC entities holding more than 85\% of all mined BTC coins so far. Such a concentration in the different aspects of the BTC system poses questions and serious doubts on its capabilities as a decentralized, non-central authority-ruled, and therefore trusted economical system. Our results are summarized as follows:

\noindent\textbf{BTC network size:} despite the fuss around BTC and crypto-currencies in general, the size of the active BTC P2P network has remained constant since early 2018, right after the main drop in BTC price.

\noindent\textbf{BTC network location:} the BTC network is mainly located in western countries, being US, Germany and France the dominant hosting countries, with more than 50\% of the active nodes. Active nodes are mainly hosted by major cloud providers in EU and US -- about 65\% of all active nodes are deployed at major cloud providers, including Hetzner, Amazon, Google, DigitalOcean, and OVH. About 3\% to 4\% of the active BTC nodes use anonymous connections, i.e., connect through Tor.

\noindent\textbf{BTC mining:} despite this western network locality, more than 65\% of the BTC blocks are mined by major Chinese mining pools, calling for potential centralization and blockchain immutability/security issues (e.g, 51\% attack~\cite{btcsec_2018}). In addition, BTC mining activity is becoming increasingly hidden (i.e., the signing authority is not revealed), calling for better approaches to identify miners.

\noindent\textbf{BTC nodes stability:} mean connection time and number of re-connections of BTC nodes significantly vary among nodes hosted at different main ASes, being Comcast nodes the most unstable ones, and Tor nodes the most dynamic ones (we use TorDNSEL to circumvent dynamic IP issues).

\noindent\textbf{BTC nodes performance:} we introduce the BTC Node Index (BNI), which reflects the goodness of a node for the BTC network; nodes hosted at OVH/Hetzner tend to rank better/worse according to BNI, and Tor nodes are the worst ranked.

\noindent\textbf{BTC coins distribution:} a very small number of BTC entities or whales hold most of the BTC coins mined so far. Even if some of these are major exchanges -- such as Kraken or Binance, it heavily questions the non-central authority control nature which BTC is supposed to eliminate -- as compared with traditional monetary systems, and makes it clear that controlling the BTC market is extremely easy for such whales, reflected in the practice by the enormous volatility shown in the BTC coin price. The top-1000 BTC entities hold more than 37\% of all BTC coins, and only 4.5\% of all BTC entities hold more than 85\% of BTC coins mined so far.

The remainder of the paper is structured as follows: In Sec.~\ref{sectionII} we present a brief overview on the most relevant blockchain and BTC concepts, required to interpret the presented results. Sec.~\ref{sectionIII} describes the conceived BTC crawler, as well as the different generated and collected datasets used in the study. The core and main findings related to the study of the BTC P2P network and mining activity are presented in Sec.~\ref{sectionIV}. The analysis of the BTC coins distribution is presented in Sec.~\ref{sectionV}. Related work on BTC analysis, and in particular regarding the P2P network perspective, is presented in Sec.~\ref{sectionVI}. Conclusions are presented in Sec.~\ref{sectionVII}.

\section{Background}\label{sectionII}
\subsection{Blockchain 101}\label{background.blockchain}
A \dfn{Blockchain} is a decentralized, distributed, and public database composed of a continuously growing list of records: the \dfn{blocks}. Those blocks are linked and secured using cryptography. In particular, each block contains ($i$) a cryptographic hash of the previous block for protecting against changes in blocks already accepted in the chain, ($ii$) a timestamp for protecting against the re-usage of a block, and ($iii$) the data. By design, a block cannot be altered retroactively without the modification of all subsequent blocks and the consensus of the network. The blockchain technology allows data to be permanently added, verified, and shared securely amongst users.

In this paper, only the problem of \textit{``public blockchains''} is considered. In such blockchains, all records are considered public and any user can take part of the mechanism to add a block to the blockchain. Therefore, all users are considered as \textit{untrusted}. A fundamental problem in this kind of environment is how to achieve \textit{consensus}, i.e., how users agree on a \textit{"correct"} shared version of the blockchain. A blockchain reaches consensus for adding a block into the chain through the computation of a moderately hard cryptographic puzzle, the \dfn{Proof of Work Consensus}.

The amount of work needed to construct the blockchain is representing a measure of its reliability. Thus, users always consider as valid the blockchain that results from the highest amount of \dfn{work} and always work on extending it. The mathematical problems that need to be solved for consensus must be representative of the amount of computational power used to solve the problem and, once the solution has been found, it must be easily verified by other users. One of the most common mathematical problems used in proof of work systems - such as Bitcoin, consists in varying a nonce that is included in the block, so that the hash of the block respects some conditions. The users that are involved in this kind of consensus process are called \dfn{miners}, and are responsible for adding blocks to the blockchain. Once a block has been mined, the miners must broadcast the information to all other users that will decide whether or not the block is correct, to later on adding it to the blockchain.

\subsection{Bitcoin 101}\label{background.bitcoin}
Blockchains can typically be used to record transactions across many computers in a decentralized and distributed way, allowing so users to have a public digital ledger. This is exactly how \dfn{Bitcoin}~\cite{bitcoin} (BTC) is using this technology. BTC is a virtual currency or \dfn{crypto-currency} relying on the blockchain technology. It allows its users to process payments without the need of any central authority such as financial institutions.

The BTC blockchain uses the Proof of Work mechanism to ensure the consensus of the users on the blocks to add to the public ledger. The cryptographic puzzle has an adjustable difficulty to compensate increasing hardware speed and varying interest in running nodes over time, as well as to ensure a stable number of block per hour. The difficulty expected by the BTC consensus protocol is updated every 2,016 blocks. The network uses timestamps stored in each block header to calculate the number of seconds elapsed between generation of the first and last of those 2,016 blocks. The ideal value is 1,209,600 seconds (i.e., two weeks). If the number of seconds is below this threshold, the expected difficulty will be increased proportionally, and vice versa.

By convention, the first transaction in a block is a special transaction that creates new coins, the so-called \emph{coinbase}, that is rewarded to the miner as incentive for mining. This is the way new BTCs are put into circulation; at the same time, there is by conception a fixed number of maximum BTCs which can be mined (i.e., 21 million BTC), thus the amount of new BTCs which can be mined decreases over time, simulating the scarcity of BTC and increasing its price, as an analogy to gold mining. Besides the mining reward, the BTC core protocol also defines the usage of \dfn{transaction fees}, which are payed by the transaction issuers to motivate miners to include their transactions in a block they are mining. Finally, it is worth noticing there exist several consensus alternatives inspired or not by the Proof of Work mechanism, such as Proof of Stake, Delegated Proof of Stake, Ripple consensus, etc.

\begin{figure}[t!]
\centering
\includegraphics[width=0.85\columnwidth]{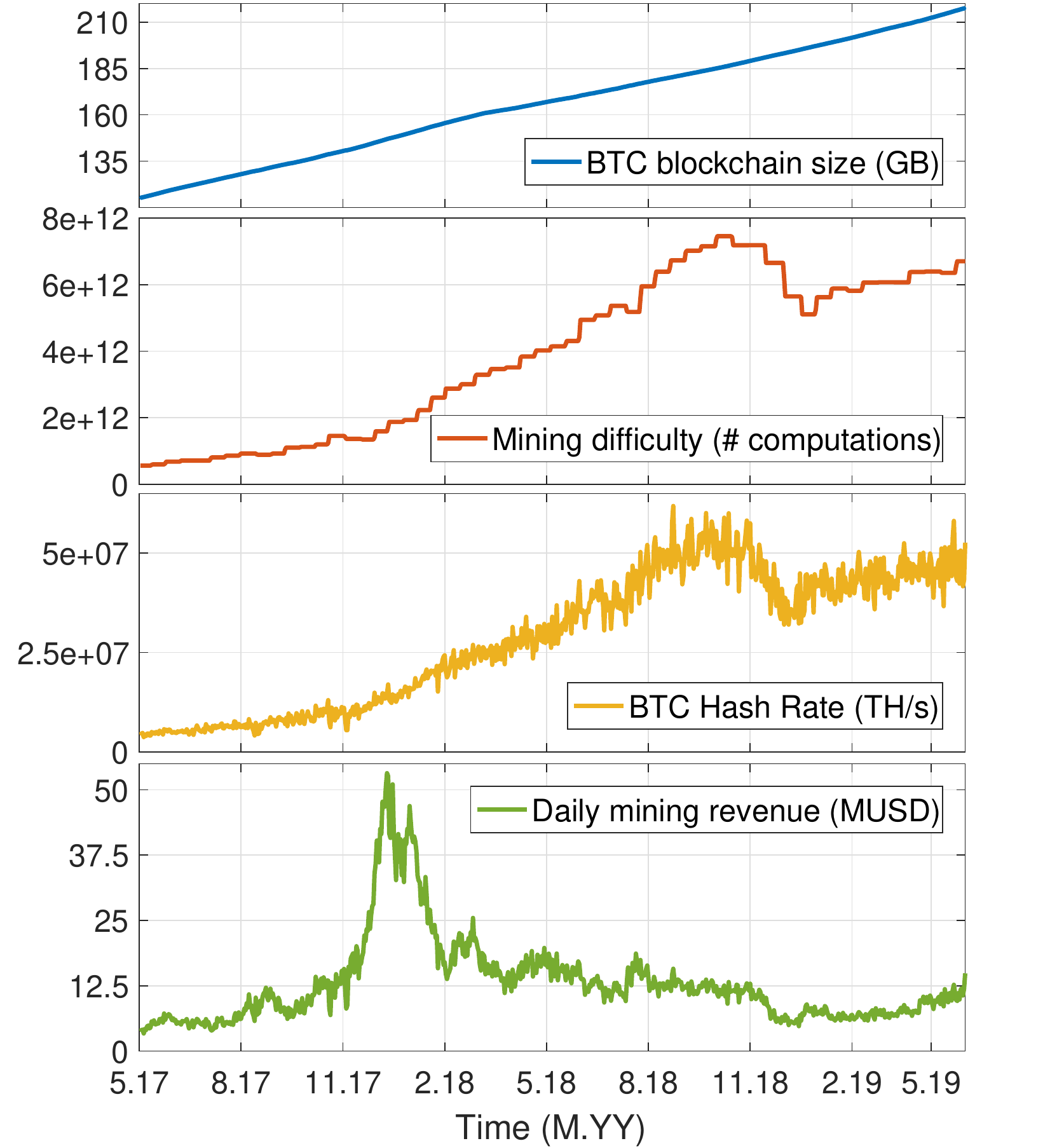}
\caption{General stats on the BTC blockchain.}
\label{fig:BTC_basics}
\end{figure}

\section{Data Collection \& Datasets}\label{sectionIII}
To discover the nodes of the BTC P2P network, we conceived a BTC crawler. This BTC crawler is a customized BTC software client which can recursively query all the BTC peers of the network, discovered by asking other nodes for the underlying IP addresses. In the start-up phase, the crawler obtains a set of \dfn{seed} or well-known IP addresses, by querying a list of predefined DNS servers. Then it starts the crawling process, by requesting to each of these seed nodes their internal list of known peers, using a protocol-defined \texttt{getaddr} message. In BTC, each node keeps a list of up to 2,500 peer addresses of tried and new connections to its neighbors~\cite{hindawi_2018}. When the crawler successfully connects to a node, it marks it as \emph{active} and requests additional meta-data -- through the protocol-defined \texttt{version} message, such as the provided services, the estimated size of the BTC blockchain -- the block height, the software version of the client, and the P2P protocol version used by the BTC node. The services provided by a BTC node~\cite{hindawi_2018} can be useful to understand which type of node it is, for example, to understand if it has the minimum capabilities to be a miner or other important role, such as a full node. Note that the list of BTC nodes does not consider most common individual end-user wallets, as these are actually provided by wallet companies which use their BTC nodes as proxies to the end-users. Once the crawling of the full P2P is completed, a set of both active and inactive BTC P2P peers is provided, along with their corresponding IP address, the meta-data and several network stats such as the latency to the node. As we show next, these measurements can be used to characterize the blockchain P2P network. The crawler heavily relies on multi-threading and multi-core platforms to improve crawling speed, currently taking less than an hour to complete a full BTC snapshot from scratch. As expected, note that the proposed BTC crawler cannot access active BTC nodes connected behind \emph{NATed} devices, but being private nodes, these are less relevant to the core activity of the BTC network, in particular offering the substrate of maintaining the DLT.

\textbf{BTC P2P network datatset:} using the BTC crawler, we take daily snapshots of the BTC network, from September 2018 to May 2019. To complement the temporal resolution of these snapshots and to extend the BTC analysis to previous dates, we rely on the APIs offered by the BTC bitnodes project (see \url{https://bitnodes.earn.com/}), which offers a similar dataset to the one generated through our BTC crawler, but with a much finer grain resolution, providing a full snapshot every five minutes, for the last 60 days. Using bitnodes, we build a dataset spanning 110 consecutive days, starting in February 2019, with a full BTC snapshot every 30 minutes. Finally, bitnodes also offers aggregated daily BTC stats over the past two years, in particular regarding BTC network size, which we also query and add to our dataset. Measurements are done from two topologically co-located vantage points in Austria and Germany, which ensures that observations from both vantage points are the same -- which we verify empirically, by comparing snapshots from both locations.

\textbf{BTC blockchain dataset:} we complement the network measurements with blockchain side statistics, accessed either through on-line APIs available at \url{https://www.btc.com} and \url{https://www.blockchain.com}, or generated directly by us, by processing the information available at the public BTC blockchain ledger of transactions. Of particular interest to our study is to recognize the miners of the BTC blocks; for doing so, we process the information available at the so called \textit{coinbase transaction}, which is a unique type of transaction, created with each new mined block: in a nutshell, when a new block is mined, the miner usually signs the associated coinbase mining-reward transaction with its name, providing as such a signature for his activity. Adding such a signature is non compulsory, so in many cases, the miner generating a new block is unknown. We collect the list of miners since early 2017 till end of 2018. Given that the BTC mining activity is usually conducted through mining pools - multiple miners share mining resources to increase their processing power, block signatures are well known for the most important mining pools -- check the list of top BTC mining pools at \url{https://gist.github.com/denpamusic/pools.json}.

\section{Unveiling the BTC IP Network}\label{sectionIV}
\subsection{BTC Blockchain}\label{sectionIV.blockchain}
We start our characterization analysis of the BTC measurements by providing a general overview on basic metrics related to the BTC blockchain. Fig. \ref{fig:BTC_basics} reports, from top to bottom, the ($i$) size of the BTC blockchain, the ($ii$) mining difficulty -- in terms of number of computations required to mine a new block, the ($iii$) aggregated BTC hash rate available at the BTC mining network, and the ($iv$) aggregated daily mining revenue, for the last two years. Each new mined BTC block is limited in size to 1 MB, and current BTC full blockchain accounts for about 218 GB. The mining difficulty is a relative measure of how difficult it is to find a new block, and the BTC hash rate is an estimated number of the hashing operations per second performed by the BTC network. Interestingly, despite the fact that the daily mining revenue has largely decreased since the end of the huge speculative BTC \emph{bubble} by end of 2017, the block mining difficulty -- and the corresponding total deployed hash power to mine new blocks, have continued increasing, with a drop by November 2018 -- the last major BTC price drop, and a subsequent recovery. Current aggregated daily mining reward is close to 15 million USD.

\noindent{\textbf{Summary:}} we can see that there is still a sustained interest and investment in the BTC mining activity despite the major drops in BTC price and reduced mining revenue, suggesting that the BTC network is on a healthy and growing uptrend.

\begin{figure}[t!]
\renewcommand{\arraystretch}{0.7}
\centering
$\begin{array}{c}
\includegraphics[width=0.85\columnwidth]{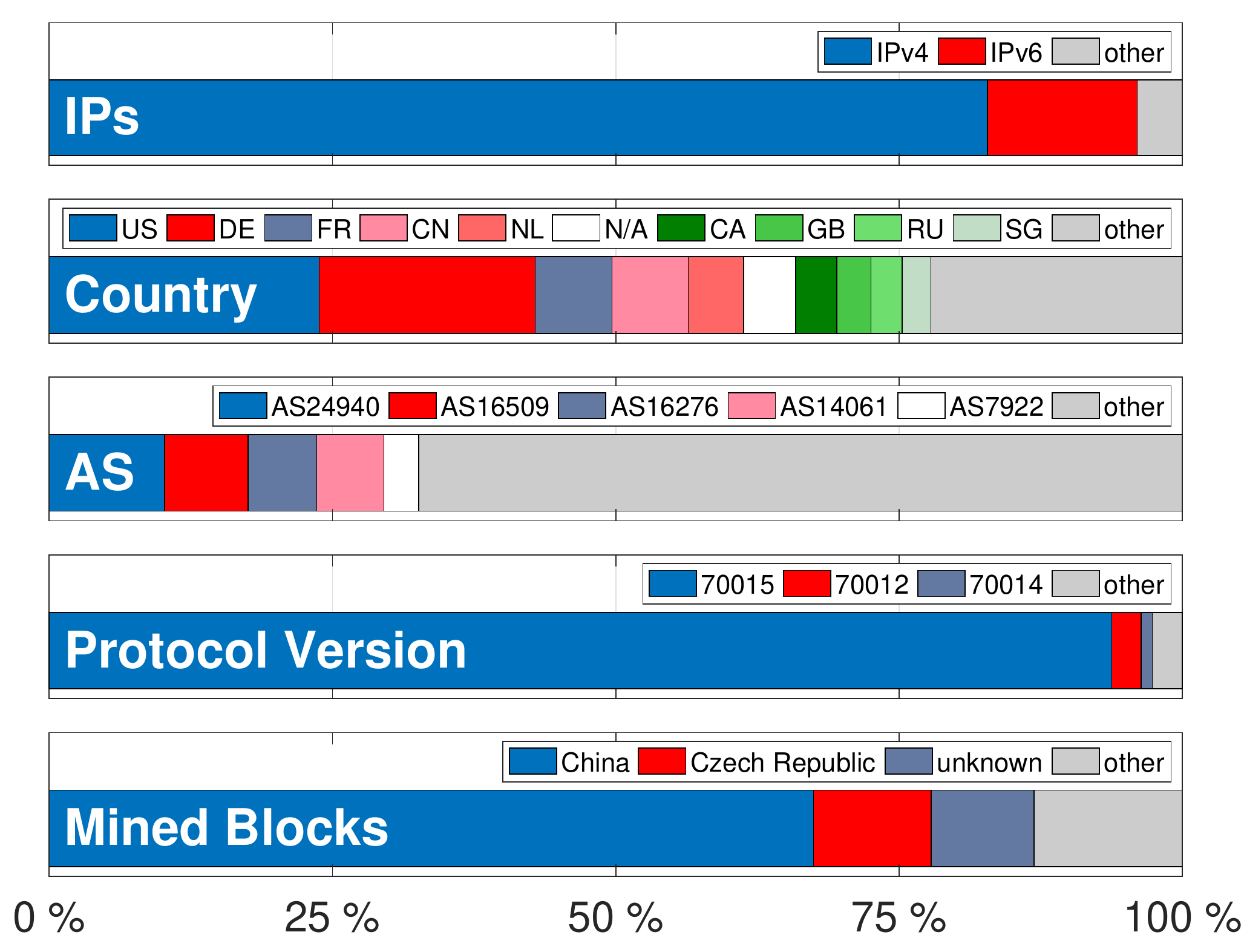}\\
\text{(a) BTC snapshot -- September 2018.}\\
\includegraphics[width=0.85\columnwidth]{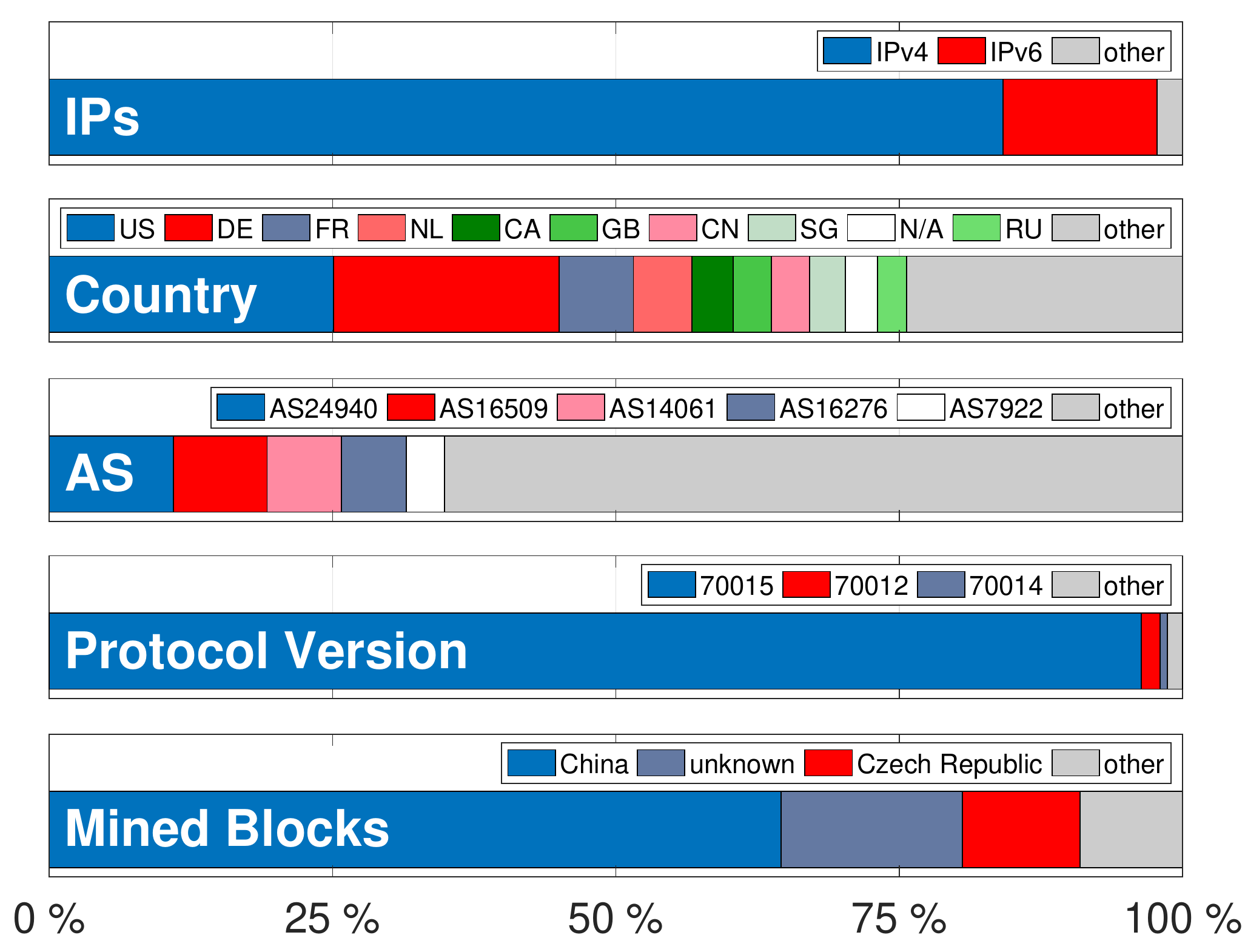}\\
\text{(b) BTC snapshot -- May 2019.}
\end{array}$
\caption{BTC nodes and network stats.}
\label{fig:btc_stats}
\end{figure}

\subsection{Crawling the BTC P2P Network}\label{sectonIV.crawling}
We describe a single snapshot of the BTC P2P network produced by our BTC crawler in September, \nth{10}, 2018. The crawler discovers more than 200,000 BTC IPs, with a total of 9,673 active nodes -- recall that this number does not reflect the actual number of BTC users/wallets, which are better captured through analysis of the BTC transactions~\cite{semantics_2016}. Fig.~\ref{fig:btc_stats}(a) presents several statistics regarding these active nodes. In terms of IP addresses, about 83\% correspond to IPv4 addresses, 13\% to IPv6, and the remaining 4\% to nodes connecting to the BTC network through anonymous connections, mostly using Tor - we rely on the public TorDNSEL service and the list of Tor exit nodes to identify these nodes.

\begin{figure}[t!]
\centering
\includegraphics[width=0.9\columnwidth]{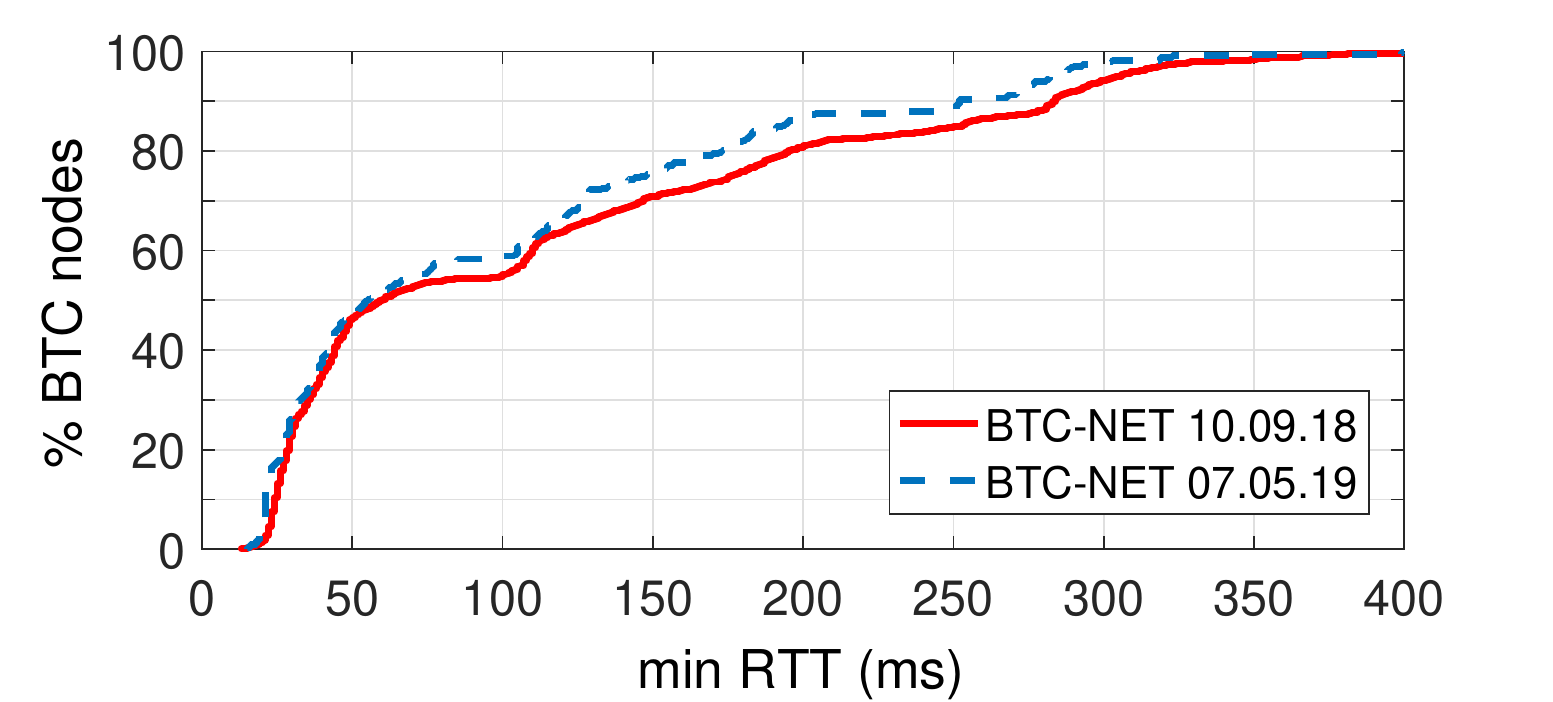}
\caption{min RTT to active BTC nodes.}
\label{fig:min_rtt_nodes}
\end{figure}

Using IP geolocation databases (both Team Cymru and MaxMind), we geolocate the IPv4 and IPv6 addresses by country, and obtain the ASes hosting them. Most of the BTC active nodes are located in the US (23.7\%) and EU -- Germany (19\%), France (6.8\%), Netherlands (4.9\%), whereas a smaller share are located in China (6.7\%) and other Asian countries (Singapore, Japan, South Korea, etc.). This is additionally confirmed by the propagation latency (min RTT) to the corresponding IP addresses~\cite{Poese_2011}, from one of the vantage points used in the data collection, located in Austria: Fig.~\ref{fig:min_rtt_nodes} shows that about 50\% of the IP addresses reside within EU (min RTT $<$ 50 ms), 30\% within North-America (100ms $<$ min RTT $<$ 200 ms), and about 10\% at very far locations, such as Eastern Asia (e.g., Japan, Korea, Hong Kong, etc.). Regarding the hosting ASes, a big share of nodes are hosted by major cloud providers such as Hetzner (AS24940), Amazon (AS16509) and OVH (AS16276). Despite such a western concentration of active nodes, it is impressive to see that the far majority of BTC mining activity actually occurs in China. As we explained before, by using the signatures provided by the miners when producing a new block, we identify the main mining pools, and observe that more than 70\% of the BTC mining activity is controlled by major Chinese mining pools -- we dig deeper into this next. The number of blocks mined by unknown or hidden miners is non-negligible, about 7\%. In terms of node services, the biggest majority of the BTC nodes keep a full copy of the full BTC DLT (93.6\%), and can therefore perform full validation of transactions and blocks; regarding P2P protocol version, while 93.8\% of the nodes use the latest version of the BTC P2P communications protocol, there are at least 2.6\% of the nodes using a very out-dated protocol, from late 2015, which can potentially result in security or performance issues.

\begin{figure}[t!]
\centering
\includegraphics[width=0.85\columnwidth]{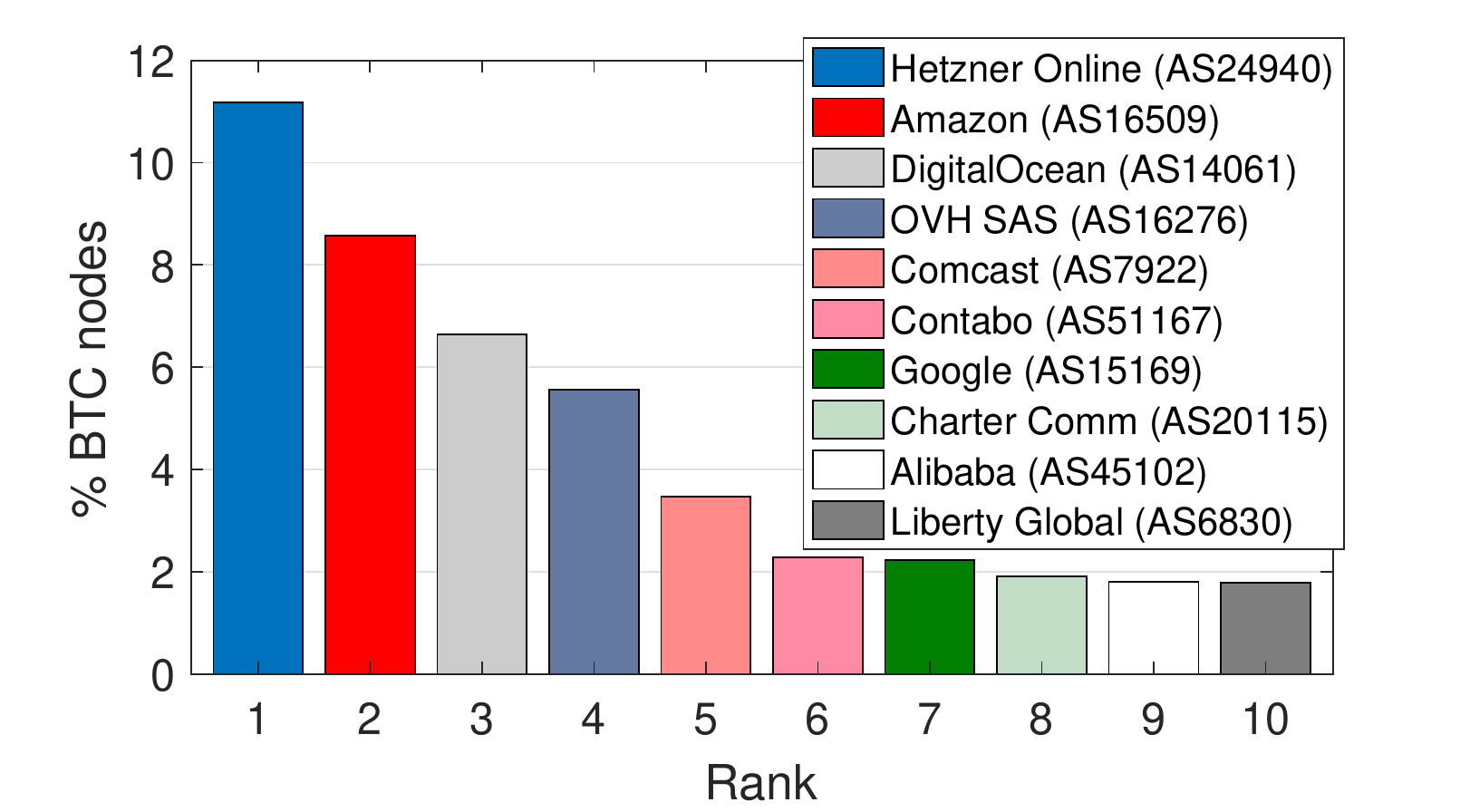}
\caption{BTC nodes hosted in top organizations.}
\label{fig:btc_top_orgs}
\end{figure}

To complement the picture, we compare these results with the latest snapshot in our dataset, recently taken in May, \nth{7}, 2019. In this new snapshot we find 9,476 active nodes. While most observations still hold, we see that the share of nodes located in China is decreasing, with a current share of 3.4\% of the total BTC active nodes. There is also a slight drop in the mining activity controlled by major Chinese pools, currently accounting for about 65\% of the BTC blocks. The share of mined blocks by unknown miners is growing, currently at about 15\% of the blocks. In terms of nodes location, Fig.~\ref{fig:min_rtt_nodes} suggests that nodes are basically the same.

Coming back to the organizations hosting most of the active BTC nodes, Fig.~\ref{fig:btc_top_orgs} reports the share of nodes hosted by the top 10 organizations by May the 7th, 2019. Not surprisingly, a very large share of BTC nodes are deployed at German cloud providers (Hetzner and Contabo), US cloud providers (DigitalOcean, Amazon, Google), Chinese cloud providers (Alibaba), as well as at major telecom players such as Comcast and Liberty Global. More than 35\% of the nodes are hosted by five major organizations, and about 65\% of all active nodes are deployed at cloud providers.

\noindent{\textbf{Summary:}} we see that the BTC P2P network infrastructure is mainly hosted at western cloud providers and ISPs, with a quite centralized infrastructure deployed at few major providers, which puts into question the decentralization nature of the BTC network. The picture looks even more critical in terms of mining activity, with about 65\% of the BTC blocks mined by major Chinese pools, with the corresponding potential security threats in terms of BTC blockchain control -- the so-called 51\% attack~\cite{btcsec_2018}.

\begin{figure}[t!]
\centering
\includegraphics[width=0.85\columnwidth]{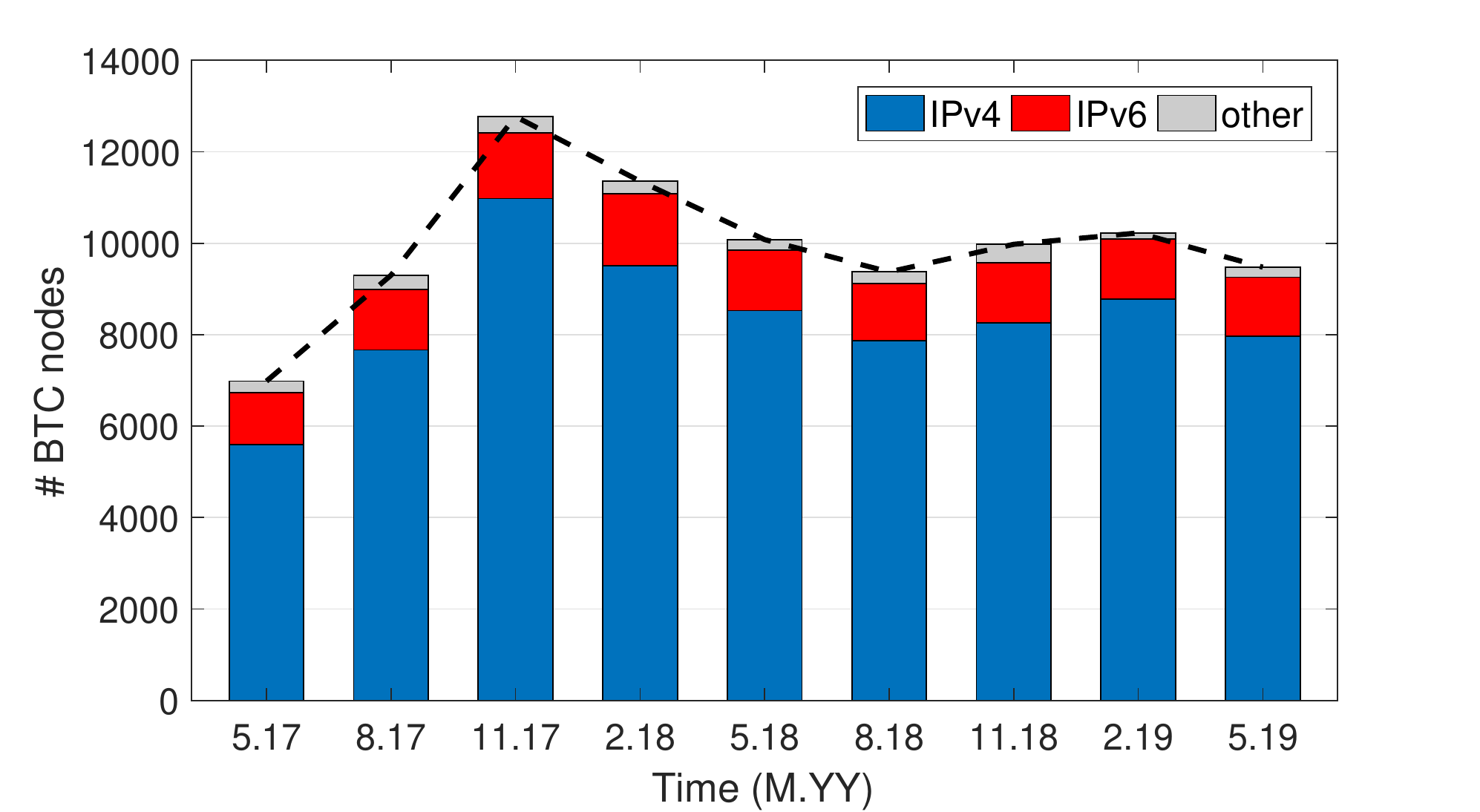}
\caption{Number of active BTC nodes along time.}
\label{fig:btc_net_size}
\end{figure}

\begin{figure}[t!]
\centering
\includegraphics[width=\columnwidth]{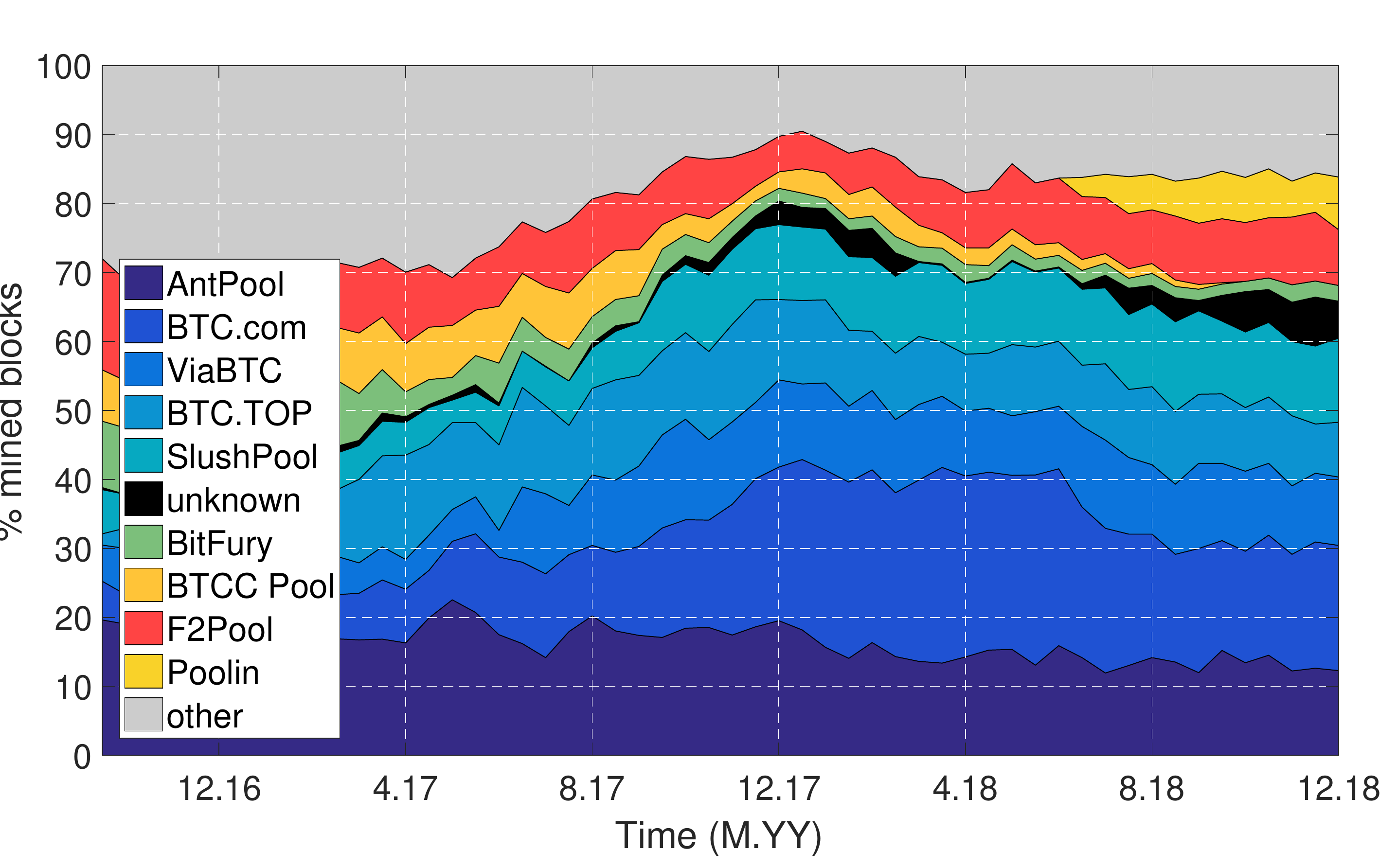}
\caption{Evolution of share of mined blocks.}
\label{fig:mined_blocks}
\end{figure}

\begin{figure}[t!]
\renewcommand{\arraystretch}{0.5}
\centering
$\begin{array}{cc}
\hspace{-5mm}\includegraphics[width=0.55\columnwidth]{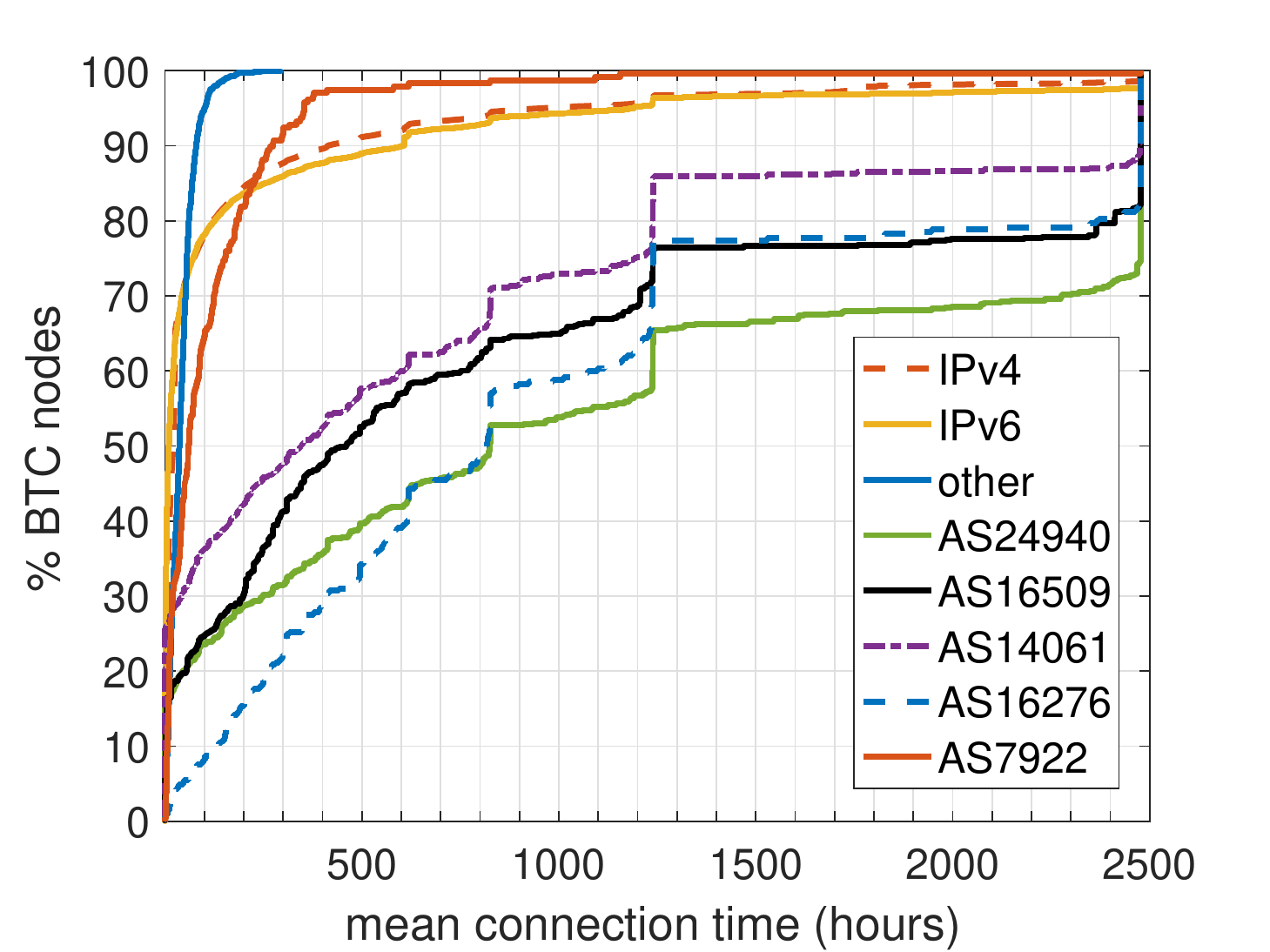} & \hspace{-5mm}\includegraphics[width=0.55\columnwidth]{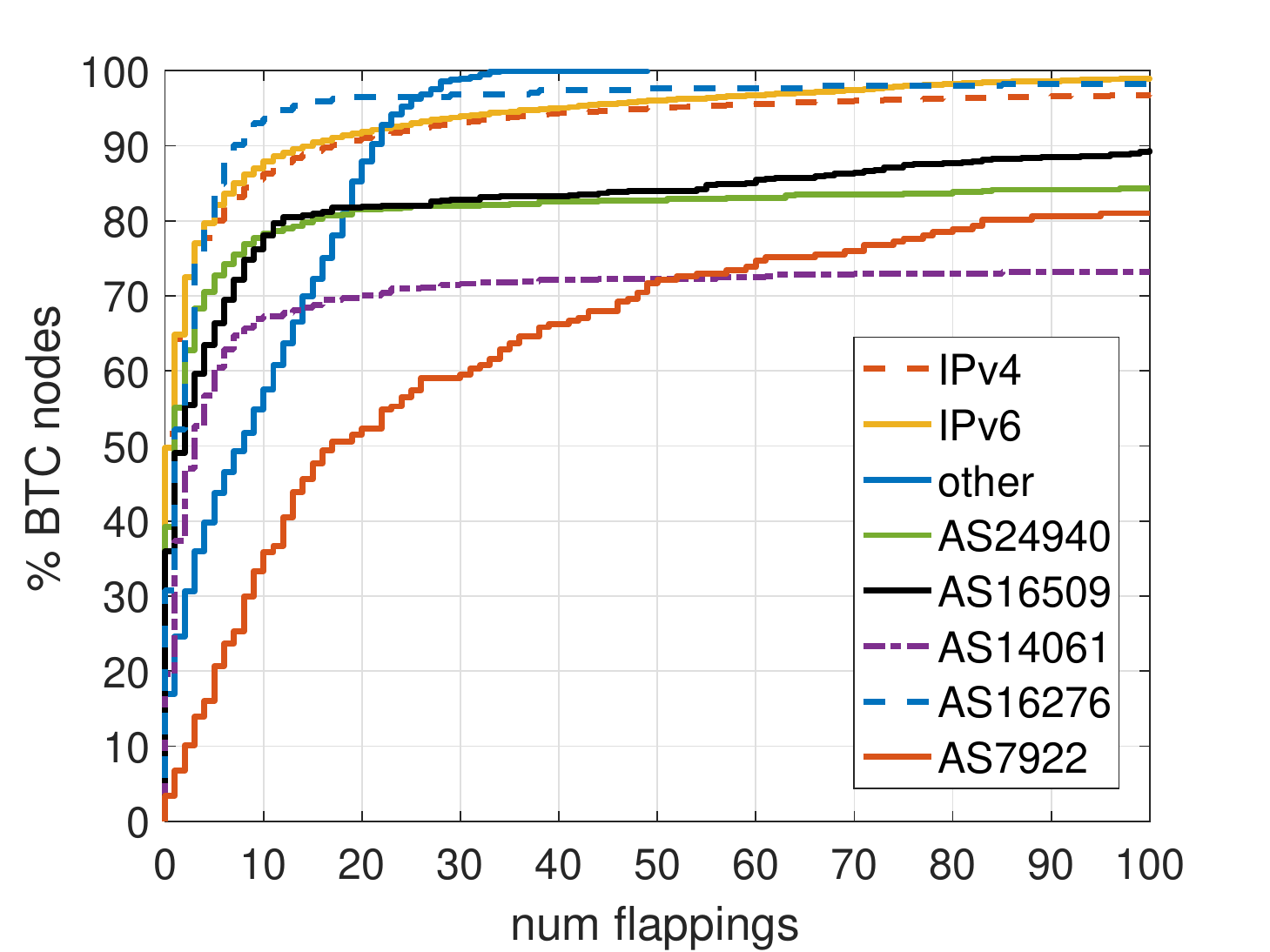}\\\vspace{-0.2mm}
\hspace{-5mm}\text{(a) Mean connection time.} & \hspace{-5mm}\text{(b) Number of flapping events.}
\end{array}$
\caption{Temporal stability of the BTC network. Results correspond to a time window of three and a half months.}
\label{fig:BTC_stability}
\end{figure}

\begin{figure*}[t!]
\renewcommand{\arraystretch}{0.7}
\centering
$\begin{array}{ccc}
\hspace{-2mm}\includegraphics[width=0.35\textwidth]{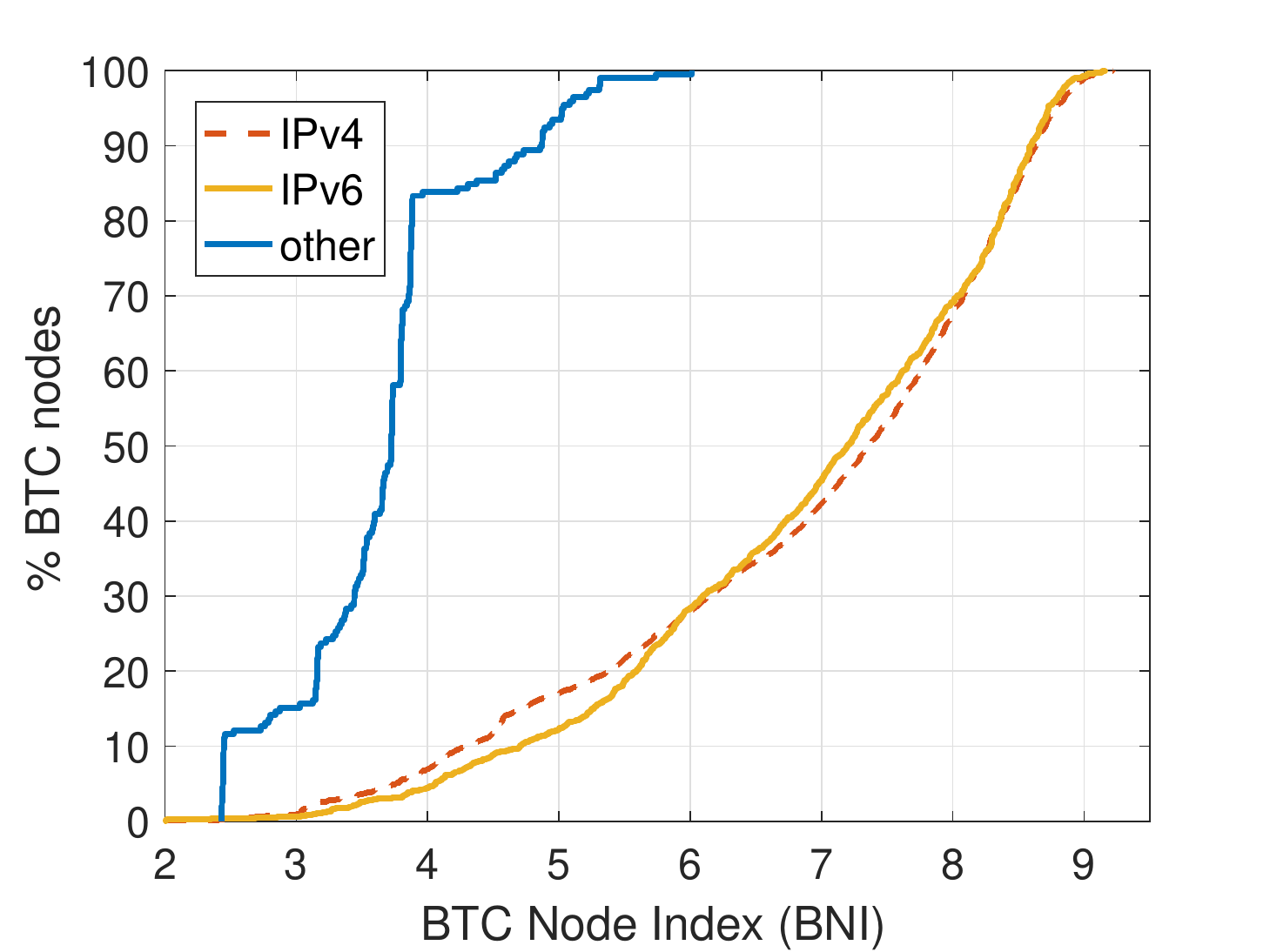} &
\hspace{-6mm}\includegraphics[width=0.35\textwidth]{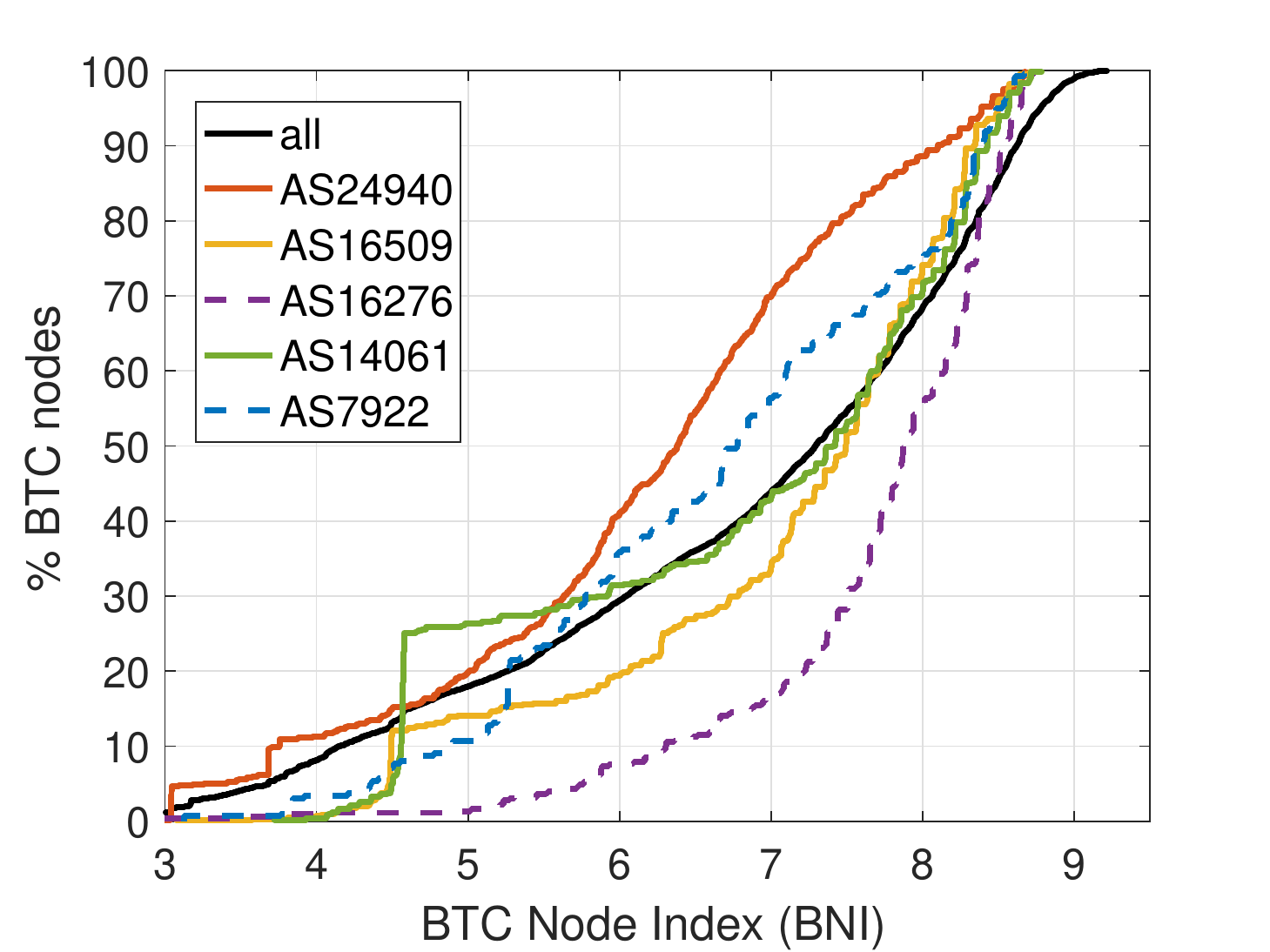} &
\hspace{-6mm}\includegraphics[width=0.35\textwidth]{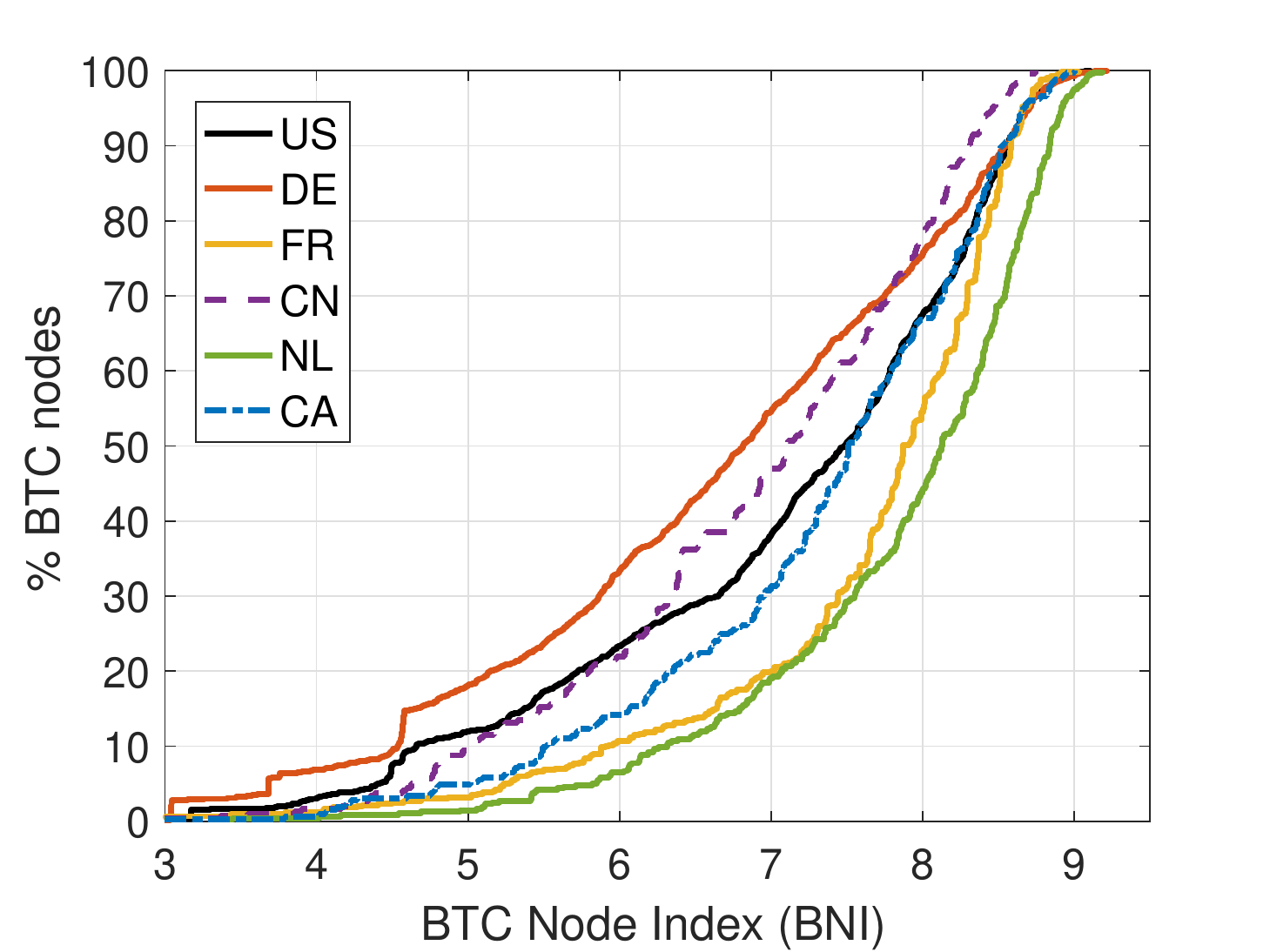}\\\vspace{-0.2mm}
\hspace{-2mm}\text{(a) BNI per network type.} & \hspace{-6mm}\text{(b) BNI per top-AS.} & \hspace{-6mm}\text{(c) BNI per top-country.}\\
\end{array}$
\caption{Bitcoin Node Index (BNI). The BNI index aggregated 10 different node-to-network metrics.}
\label{fig:BNI}
\end{figure*}

\subsection{Temporal Analysis}\label{sectionIV.temporal}
We now move on to a temporal characterization of the BTC network. We focus the analysis on two specific metrics of interest: the size of the BTC P2P network, and the share of BTC blocks' mining. Fig.~\ref{fig:btc_net_size} reports the number of active BTC nodes over the past two years, splitting by network type. While the number of active nodes grew significantly in 2017 -- almost doubling in less than 6 months, till the end of the bubble, we see that the size of the BTC active network has remained almost constant during the past year, with about 10,000 nodes daily active. This suggests that the number of users interested in running BTC (full) nodes is not growing, and thus, the underlying P2P network is rather stable in terms of new members.

Fig.~\ref{fig:mined_blocks} plots the temporal evolution of the share of mined blocks, during the past 2 years, till end of 2018. Here we split the shares by mining pool signature. The top mining pools by end of 2018 in terms of blocks is lead by major Chinese companies such as AntPool (12.3\%), BTC.com (18.2\%), ViaBTC (9.9\%), BTC.top (8\%), F2Pool (8\%), and Poolin (7.6\%), keeping a similar dominance along time. Other non-Chinese pools include SlushPool (12.2\%) in Czech Republic and BitFury (2.3\%) in Georgia. As we saw in the snapshots analysis, the share of unknown mined blocks is growing, at around 6\% to 7\% by end of 2018, and close to 15\% these days. It is interesting to observe how certain mining pools such as BTCC Pool, located in Hong Kong, left the market fully by end of 2018, with some other new pools taking over, such as Poolin.

\noindent{\textbf{Summary:}} the temporal analysis reveals that the size of the BTC P2P network has remained constant over the past year, and that the Chinese dominance in terms of mining activity holds along time, with a very centralized structure. The main four Chinese mining pools have covered more than 50\% of the mining activity since roughly mid 2017, which is a serious security threat for the integrity and reliability of the whole BTC network.

In terms of node stability, and being the BTC network a P2P network, we study the fluctuation of active nodes along time in terms of mean connection time and number of flapping events, the latter corresponding to non-contiguous time periods of activity, i.e., an active node which disconnects and becomes active at a posterior time. The study is conducted on top of the BTC snapshots collected every 30 minutes for 110 consecutive days, using bitnodes. Fig.~\ref{fig:BTC_stability} depicts ($a$) the mean connection time and ($b$) the number of flapping events for all BTC active nodes over the corresponding period. While the usage of dynamic IP addresses would naturally result in a misleading interpretation, Fig.~\ref{fig:BTC_stability}(a) shows that both IPv4 and IPv6 nodes have an almost identical distribution in terms of mean connection time; if we take into account that IPv6 addresses tend to be static, and in particular in major cloud providers~\cite{Chen_2015} -- due to the large availability of them, we can conclude that the provided results are quite close to the real situation. Mean connection time and number of re-connections of BTC nodes significantly varies among nodes hosted at different main ASes, being Comcast (AS7922) nodes the most unstable ones, and Tor nodes the most dynamic ones -- note that we use TorDNSEL to circumvent dynamic IP issues.

\subsection{BTC Node Performance}\label{sectionIV.btcperf}
The last part of the network study is devoted to the analysis of the performance of each BTC P2P node. To do so, we define the \dfn{BTC Node Index} (BNI)
The BNI ranges from 0 to 10, with 10 representing the best fitted node for the BTC network, and 0 the worst. BNI is computed as an average of 10 different node-to-network metrics, which basically reflect how similar is a node to the majority or most common node, as well as how good is this node connected and synchronized to the network. These 10 metrics include: ($i$) the protocol version index = 1/$r$, being $r$ the rank of the node's protocol version -- e.g., it is equal to 1 if the version is 70015 (cfr. Fig.~\ref{fig:btc_stats}), ($ii$) the service index -- reflects how similar are the offered capabilities by this node, ($iii$) the port index -- 1 if uses the default BTC port, 8333, ($iv$) the block height index -- how well synchronized is this node to the rest in terms of knowledge of the BTC network size, ($v$) the ASN index = ln((1/$n$) $\times$ $N$)/ln($N$), with $n$ and $N$ the number of nodes in the same AS as this node and the size of the network, as well as five additional metrics reflecting the daily/weekly latency performance variation -- variations in min RTT above a threshold, cfr. Fig.~\ref{fig:min_rtt_nodes}, the node uptime, and its availability. Note that, to avoid the bias introduced by using latency measurements from a fix vantage point, the latency performance variation threshold is computed independently for each node, as a fix percentage of the moving average of past measurements. In a nutshell, these latency performance metrics flag major latency increases for active nodes.

Fig.~\ref{fig:BNI} reports the BNI of the network for the last week of the 30'-grained dataset, splitting by ($a$) network type, ($b$) top ASes, and ($c$) top countries. There is no significant difference between IPv4 and IPv6 nodes, but Tor nodes are significantly worse ranked, mainly due to their mismatch to the majority of the nodes in terms of node properties. In terms of ASes, nodes hosted by OVH and Amazon outperform those hosted by other main ASes, with a significant difference between OVH and Hetzner, which performs the worst. Finally, nodes hosted at the Netherlands and France systematically rank higher than nodes hosted in US and China, among others. Due to space limitation, we leave a deeper analysis of the components of the BNI index as future work.

\begin{figure}[t!]
\centering
\includegraphics[width=0.75\columnwidth]{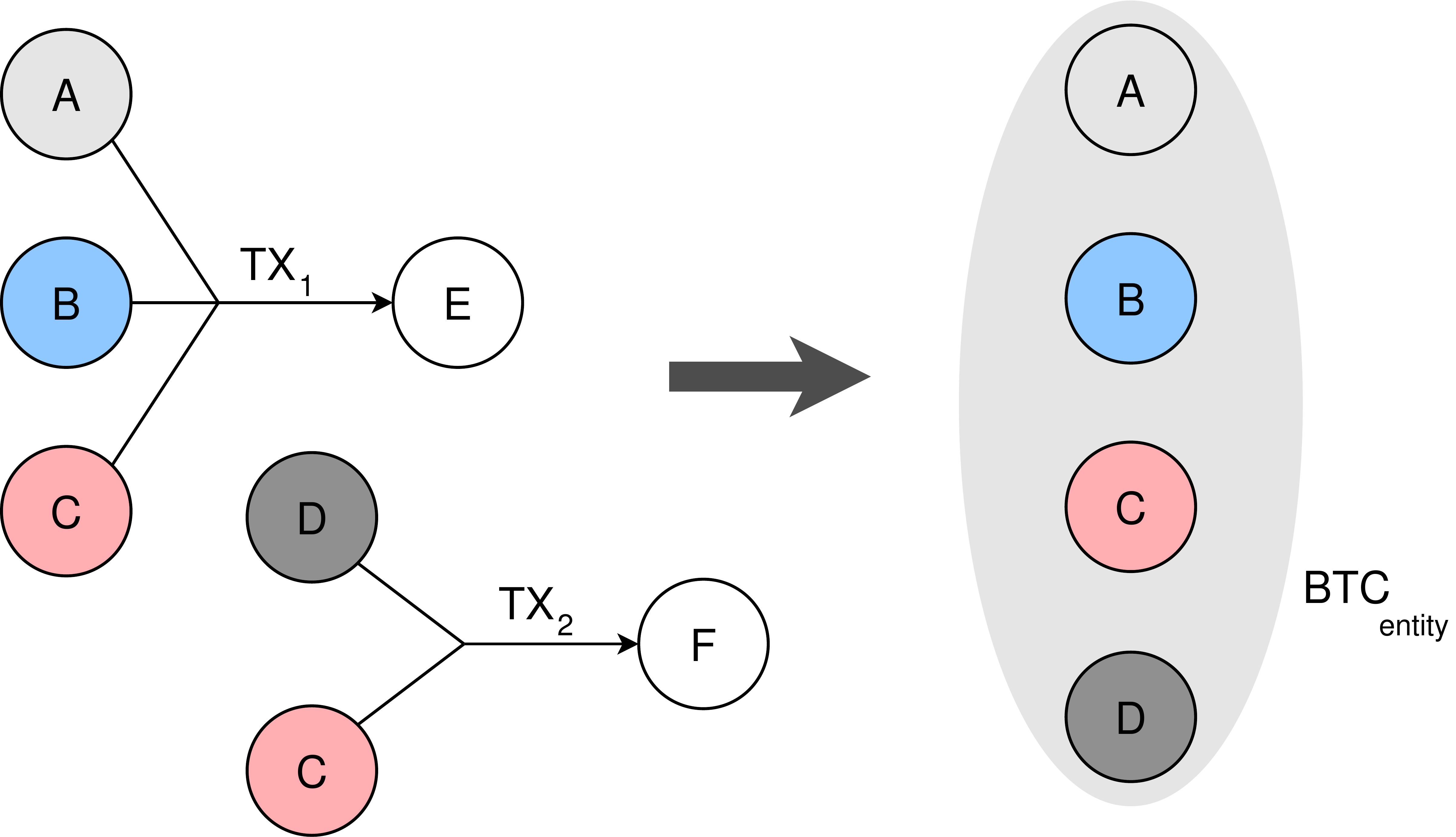}
\caption{Clustering heuristic to identify BTC entities.}
\label{fig:BTC_clustering}
\end{figure}

\section{BTC Coins -- Unveiling the Big Whales}\label{sectionV}
To conclude the paper, we focus now on the analysis of the distribution of BTC coins among independent BTC entities. We refer to the term \dfn{entity} as either single BTC addresses, or groups of BTC addresses controlled by the same actor. For example, an exchange would normally have control over a large number of BTC addresses -- potentially representing the customers or groups of them, and even if the BTC coins associated to those addresses do not belong to the exchange, in the practice it has full control over the coins, acting as a sort of centralized bank holding the coins of its customers.

To identify such entities, we can partition the BTC addresses space into clusters -- which are likely controlled by the same actor, using well-known multiple-input clustering heuristics~\cite{semantics_2016,clustering_CBT17}. Fig.~\ref{fig:BTC_clustering} explains the basic idea behind this clustering approach. The underlying assumption is that if multiple BTC addresses -- let us say A, B, and C, are used as input in the same transaction $\text{TX}_\text{1}$, while at least one of these addresses along with another address -- let us say C and D, are observed as input in another transaction $\text{TX}_\text{2}$, then the four addresses A, B, C, and D must somehow be controlled by the same actor, who performed both transactions through the corresponding private keys. Of course, this heuristic fails when CoinJoin transactions (see \url{https://en:bitcoin:it/wiki/CoinJoin}) are used, because these combine payments from different spenders that do not necessarily represent one single entity. Aware of this problem, we filter those transactions before applying the clustering heuristics.

\begin{figure}[t!]
\renewcommand{\arraystretch}{0.7}
\centering
$\begin{array}{c}
\includegraphics[width=0.85\columnwidth]{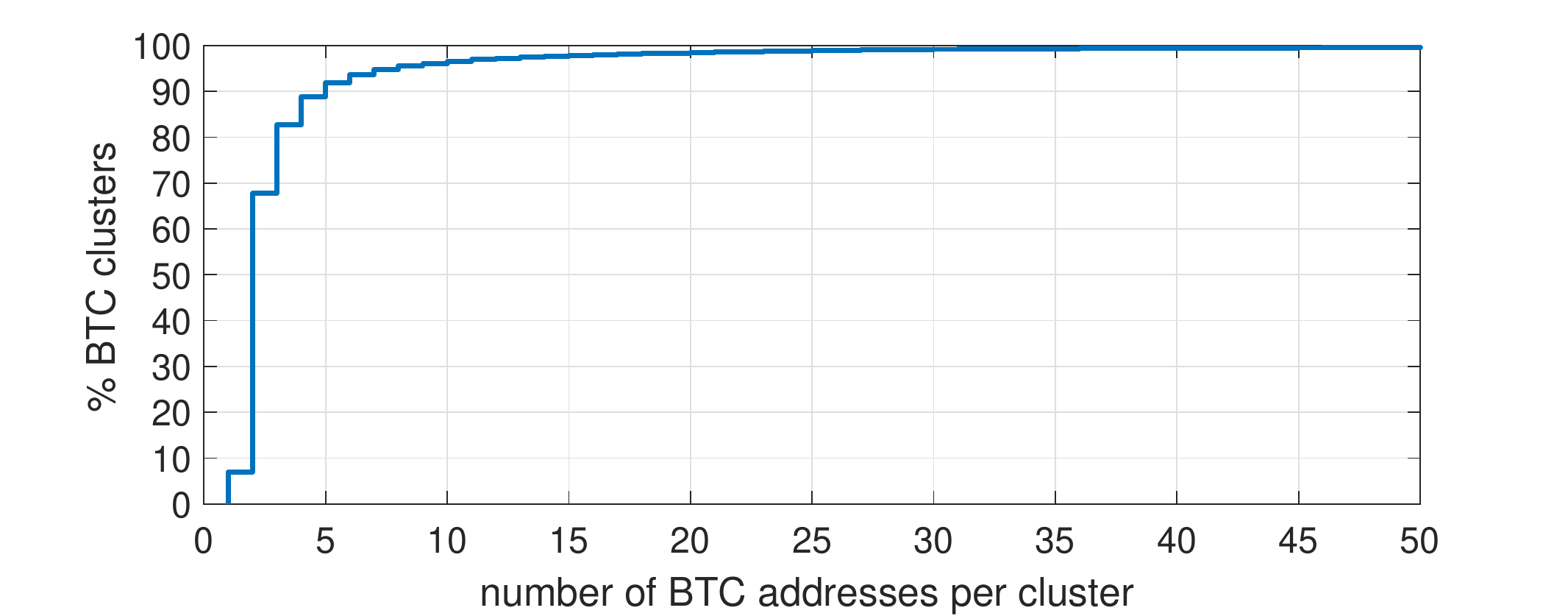}\\\vspace{-0.2mm}
\text{(a) Cluster (BTC entities) size.}\\
\includegraphics[width=0.85\columnwidth]{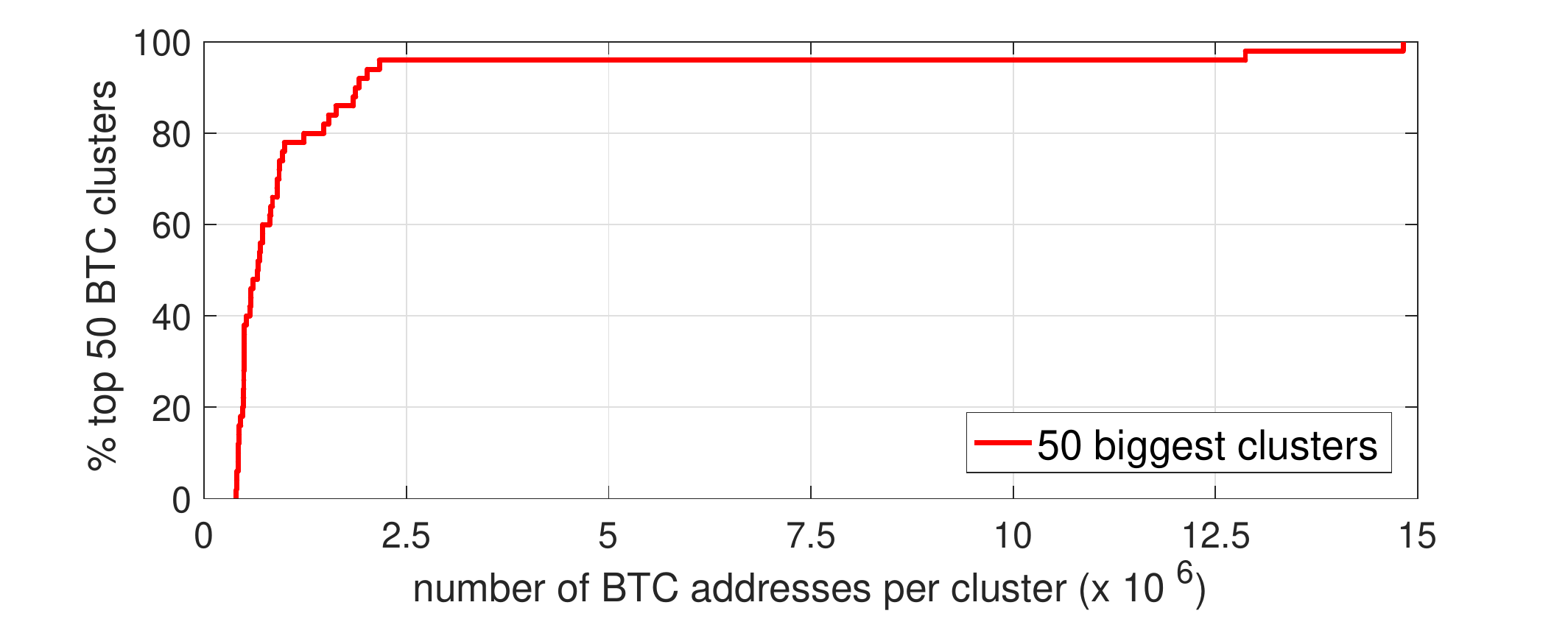}\\\vspace{-0.2mm}
\text{(a) Top-50 cluster (BTC entities) size.}\\
\end{array}$
\caption{Distribution of BTC cluster size.}
\label{fig:btc_clusters}
\end{figure}

By clustering the BTC ledger available in March, \nth{20}, 2019, we identify a total of almost 51 million clusters or entities. Fig.~\ref{fig:btc_clusters} depicts the distribution of the entity size -- number of BTC addresses per cluster, for all the identified clusters. As expected, the largest majority of entities have a small number of BTC addresses associated to them, with almost 70\% of them having 1 or 2 BTC addresses. Still, as depicted in Fig.~\ref{fig:btc_clusters}(b), the top-50 biggest BTC entities in terms of number of addresses can be extremely big, having hundreds of thousands and even millions of BTC addresses associated. While not being verifiable without having attribution data, these statistics suggest that such big clusters represent exchange services or wallet providers.

\begin{figure}[t!]
\centering
\includegraphics[width=0.85\columnwidth]{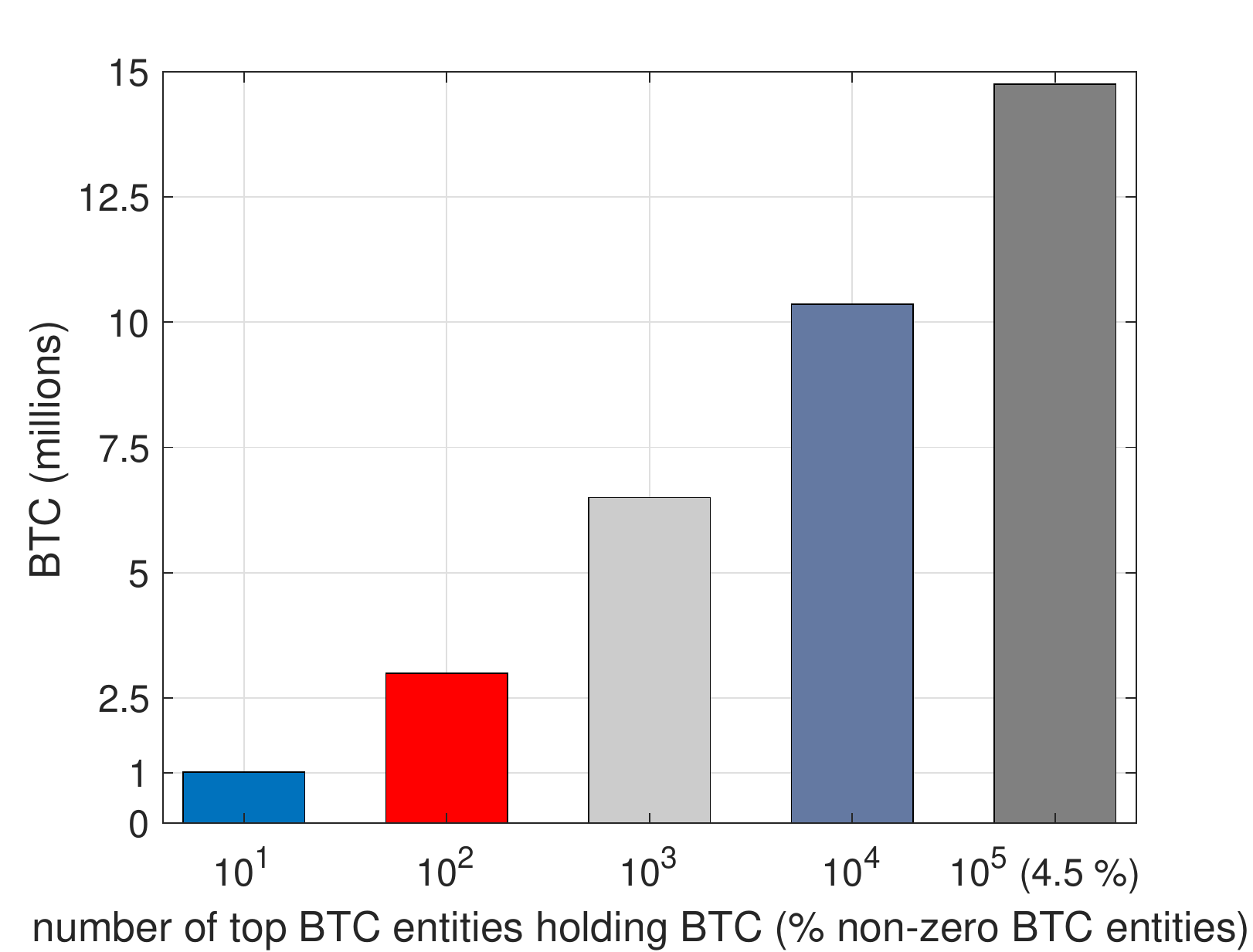}
\caption{BTC distribution for non-zero-balance entities.}
\label{fig:btc_hodlers}
\end{figure}

\begin{figure}[t!]
\centering
\includegraphics[width=0.65\columnwidth]{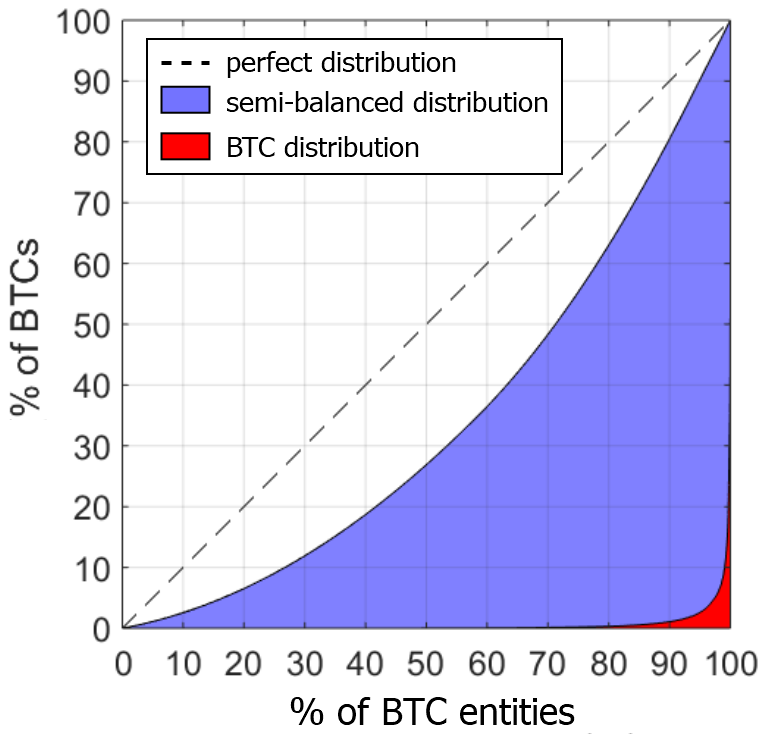}
\caption{Lorenz curve -- distribution of BTC. The Gini index for BTC in March 2019 is 0.9888.}
\label{fig:btc_gini}
\end{figure}

We now proceed to verify the balance -- in terms of BTC coins, available by March, \nth{20}, 2019, at each of these BTC entities. To get the balance of a BTC entity, we simply subtract the total outcomes from the total incomes, for all the BTC addresses associated to this entity. As expected, the largest majority of BTC entities have a zero balance, with only 2.2 million entities showing a BTC balance above 0. The interesting part comes when analyzing the BTC distribution among these non-zero-balance entities. Fig.~\ref{fig:btc_hodlers} shows the total number of BTC coins available on the top BTC entities -- top in terms of BTC coins balance. The first 10 BTC entities hold about 1 million BTC coins (1 M-BTC), which corresponds to about 5.7\% of the total 17.6 M-BTC mined till March the 20th, 2019. The top 3 BTC entities, as well as the \nth{5} and \nth{9}, correspond to individual BTC addresses, the \nth{4} and \nth{8} top entities are huge clusters of more than 100,000 and 575,000 addresses respectively - potentially popular exchanges, as we observed from publicly available attribution data, and the rest are smaller clusters of not more than 10 addresses. The top 100,000 BTC entities, which correspond roughly to 4.5\% of the non-zero-balance entities, hold almost 15 M-BTC, which is about 85\% of the total BTC coins availability by March, \nth{20}. Finally, Fig.~\ref{fig:btc_gini} plots the so-called Lorenz curve, which is a graphical representation of the distribution of a certain asset, typically used in economics to study the distribution of wealth in society. A diagonal line (see the dashed line) represents perfect distribution or equality among all entities; in blue we plot the Lorenz curve for a typically semi-balanced wealth distribution country, in this case the example is a EU country, and in red we plot the Lorenz curve for BTC. The associated Gini index -- the area between the diagonal and the Lorenz curve, i.e., 0 under perfect equality and 1 under perfect inequality, is 0.3129 for the example EU country, and 0.9888 for BTC, showing the highly concentrated BTC distribution.

\section{Related Work}\label{sectionVI}
Previous papers have studied the BTC blockchain, mainly in terms of executed transactions, through the analysis of the publicly available DLT. For example, papers such as Meiklejohn et al.~\cite{imc_2013} and Haslhofer et al.~\cite{semantics_2016} focus on the BTC transactions as observed at the BTC DLT, Tapsell et al.~\cite{btcsec_2018} study the security of the BTC P2P network, Bowden et al.~\cite{block_arrivals_2018} analyze the temporal generation of BTC blocks, O'Dwyer and Malone~\cite{btc_energy} focus on the energy footprint of BTC mining, etc.

Other papers study the P2P network topology and characteristics of BTC and other popular blockchains~\cite{imc_2018,imc_2018_webmine,bitcoin_topo_2016,coinscope_2015}. However, the continuous evolution of the DLT technology and the potential security issues linked to unveiled P2P topologies~\cite{gnutella_2002,gnutella_2005} result in constant updates of the underlying protocols, turning some of the previous proposals no longer applicable. In particular, back in 2015, Miller et al.~\cite{coinscope_2015} proposed a comprehensive technique to discover P2P links in the BTC network and identify topologically-influential nodes, relying on the analysis of the broadcast messages over the network. However, the proposed technique is no longer applicable to current BTC protocol, which has been updated to remove relevant timing information used by Miller et al.~\cite{coinscope_2015}. In particular, the techniques introduced by Miller et al.~\cite{coinscope_2015} took advantage of the two-hours penalty applied to received address messages from connected peers, which could be exploited to reconstruct the topology. However, this two-hour penalty was removed from the Bitcoin Core nodes after 0.10.1 release, reducing the fingerprint left by address messages, and therefore, making the approach no longer useful to infer the topology of the BTC network.

Closer to our work, Gencer et al.~\cite{Decentralization_BTC} and Delgado-Segura et al.~\cite{BTC_measurements} present different yet complementary approaches to infer relevant networking characteristics of the BTC P2P network. Delgado-Segura et al.~\cite{BTC_measurements} focus is on presenting a particular technique to infer the topology of the publicly reachable BTC network using the so-called \emph{orphan transactions} -- transactions that arrive out of order, but not on characterizing the actual main BTC network -- presented results consist of validating the proposed techniques over the testnet network. Similar to our work, Gencer et al.~\cite{Decentralization_BTC} do perform a characterization of the BTC P2P network, but using a different platform and methodologies to infer the nodes. Their results suggest that around 56\% of Bitcoin nodes are run in datacenters, which we confirm in this study, with a similar share, of actually 65\%. Compared to these papers, our study goes deeper in terms of performance metrics, longitudinal (i.e., temporal) characterization of the BTC network, as well as focusing on the BTC coins distribution.

\section{Conclusions}\label{sectionVII}

We have characterized the BTC P2P network, analyzing its nodes from a purely network measurements-based approach. By crawling and locating the active BTC nodes, as well as by studying the associated BTC mining activity, we conclude that the BTC network has remained rather fixed in terms of size over the past year, and that it is far from the so-much advertised decentralized and non-controlled system it should be, with a major dominance of western cloud providers and ISPs in terms of hosting most of the active nodes, and a highly dangerously concentrated mining activity at major BTC mining pools in China. Better BTC nodes are usually located at EU-centric countries, outperforming node performance in US and China. In terms of BTC coins distribution, the picture looks even worse, with most of coins under the control of a very small number of big whales.

\section*{Acknowledgments}

The research leading to these results has been partially funded by the Vienna Science and Technology Fund (WWTF) through project ICT15-129, ``Big-DAMA''. Work on this topic is supported inter alia by the European Union's Horizon 2020 research and innovation programme under grant agreement No. 740558 (TITANIUM) and the Austrian FFG's KIRAS programme under project VIRTCRIME (No. 860672).

\balance
\bibliographystyle{IEEEtran}
\bibliography{Bibliography}

\begin{thebibliography}{10}
\providecommand{\url}[1]{#1}
\csname url@samestyle\endcsname
\providecommand{\newblock}{\relax}
\providecommand{\bibinfo}[2]{#2}
\providecommand{\BIBentrySTDinterwordspacing}{\spaceskip=0pt\relax}
\providecommand{\BIBentryALTinterwordstretchfactor}{4}
\providecommand{\BIBentryALTinterwordspacing}{\spaceskip=\fontdimen2\font plus
\BIBentryALTinterwordstretchfactor\fontdimen3\font minus
  \fontdimen4\font\relax}
\providecommand{\BIBforeignlanguage}[2]{{%
\expandafter\ifx\csname l@#1\endcsname\relax
\typeout{** WARNING: IEEEtran.bst: No hyphenation pattern has been}%
\typeout{** loaded for the language `#1'. Using the pattern for}%
\typeout{** the default language instead.}%
\else
\language=\csname l@#1\endcsname
\fi
#2}}
\providecommand{\BIBdecl}{\relax}
\BIBdecl

\bibitem{hindawi_2018}
S.~Delgado-Segura, C.~P{\'e}rez-Sol{\`a}, J.~Herrera-Joancomarti,
  G.~Navarro-Arribas, and J.~Borrell, ``Cryptocurrency networks: A new {P2P}
  paradigm,'' \emph{Mobile Information Systems}, vol. 2018, March 2018.

\bibitem{semantics_2016}
B.~Haslhofer, R.~Karl, and E.~Filtz, ``O bitcoin where art thou? insight into
  large-scale transaction graphs,'' in \emph{Proc. SEMANTICS}, September 2016.

\bibitem{clustering_CBT17}
A.~Judmayer, A.~Zamyatin, N.~Stifter, A.~G. Voyiatzis, and E.~R. Weippl,
  ``Merged mining: Curse or cure?'' in \emph{Proc. International Workshop
  Cryptocurrencies and Blockchain Technology ({CBT})}, September 2017.

\bibitem{gnutella_2002}
M.~Ripeanu, I.~Forster, and A.~Iamnitchi, ``Mapping the {G}nutella network,''
  \emph{{IEEE} Internet Computing}, vol.~6, no.~1, pp. 50--57, January 2002.

\bibitem{gnutella_2005}
D.~Stutzbach and R.~Rejaie, ``Capturing accurate snapshots of the gnutella
  network,'' in \emph{Proc. {IEEE} {INFOCOM}}, April 2005.

\bibitem{btcsec_2018}
J.~Tapsell, G.~N. Akram, and K.~Markantonakis, ``An evaluation of the security
  of the bitcoin peer-to-peer network,'' in \emph{Proc. {IEEE} International
  Conference on Blockchain}, July/August 2018.

\bibitem{bitcoin}
S.~Nakamoto, ``Bitcoin: A peer-to-peer electronic cash system,'' Tech. Rep.,
  2018.

\bibitem{Poese_2011}
I.~Poese, S.~Uhlig, M.~A. Kaafar, B.~Donnet, and B.~Gueye, ``{IP} geolocation
  databases: Unreliable?'' \emph{{ACM} {SIGCOMM} Computer Communication
  Review}, vol.~41, no.~2, pp. 53--56, April 2011.

\bibitem{Chen_2015}
F.~Chen, R.~K. Sitaraman, and M.~Torres, ``End-user mapping: Next generation
  request routing for content delivery,'' in \emph{Proc. {ACM} {SIGCOMM}},
  August 2015.

\bibitem{imc_2013}
S.~Meiklejohn, P.~M., G.~Jordan, K.~Levchenko, D.~McCoy, G.~M. Voelker, and
  S.~Savage, ``A fistful of bitcoins: Characterizing payments among men with no
  names,'' in \emph{Proc. {ACM} Internet Measurement Conference ({IMC})},
  October 2013.

\bibitem{block_arrivals_2018}
R.~Bowden, H.~P. Keeler, A.~E. Krzesinski, and P.~G. Taylor, ``Block arrivals
  in the bitcoin blockchain,'' ar{X}iv, cs.{CR} 1801.07447, January 2018.

\bibitem{btc_energy}
K.~J. {O'Dwyer} and D.~Malone, ``Bitcoin mining and its energy footprint,'' in
  \emph{Proc. IET Irish Signals Systems Conference and China-Ireland
  International Conference on Information and Communications Technologies
  ({ISSC/CIICT})}, June 2014.

\bibitem{imc_2018}
S.~K. Kim, Z.~Ma, S.~Murali, J.~Mason, A.~Miller, and M.~Bailey, ``Measuring
  {E}thereum network peers,'' in \emph{Proc. {ACM} Internet Measurement
  Conference ({IMC})}, October/November 2018.

\bibitem{imc_2018_webmine}
J.~R{\"u}th, T.~Zimmermann, K.~Wolsing, and O.~Hohlfeld, ``Digging into
  browser-based crypto mining,'' in \emph{Proc. {ACM} Internet Measurement
  Conference ({IMC})}, October/November 2018.

\bibitem{bitcoin_topo_2016}
T.~Neudecker, P.~Andelfinger, and H.~Hartensetin, ``Timing analysis for
  inferring the topology of the bitcoin peer-to-peer network,'' in \emph{Proc.
  IEEE International Conference on Advanced and Trusted Computing ({ATC})},
  July 2016.

\bibitem{coinscope_2015}
A.~Miller, J.~Litton, A.~Pachulski, N.~Gupta, D.~Levin, N.~Spring, and
  B.~Bhattacharjee, ``Discovering bitcoin's public topology and influential
  nodes,'' May 2015, see
  \url{https://www.cs.umd.edu/projects/coinscope/coinscope.pdf}.

\bibitem{Decentralization_BTC}
A.~E. Gencer, S.~Basu, I.~Eyal, R.~Van~Renesse, and E.~G. Sirer,
  ``Decentralization in bitcoin and ethereum networks,'' in \emph{Proc.
  International Conference on Financial Cryptography and Data Security ({FC})},
  February/March 2018.

\bibitem{BTC_measurements}
S.~Delgado{-}Segura, S.~Bakshi, C.~P{\'{e}}rez{-}Sol{\`{a}}, J.~Litton,
  A.~Pachulski, A.~Miller, and B.~Bhattacharjee, ``Txprobe: Discovering
  bitcoin's network topology using orphan transactions,'' in \emph{Proc.
  International Conference on Financial Cryptography and Data Security ({FC})},
  February 2019.

\end{thebibliography}
\end{document}